# Robo-AO: autonomous and replicable laser-adaptive-optics and science system

C. Baranec*[a], R. Riddle[a], A. N. Ramaprakash[b], N. Law[c], S. Tendulkar[a], S. Kulkarni[a], R. Dekany[a], K. Bui[a], J. Davis[a], M. Burse[b], H. Das[b], S. Hildebrandt[a], S. Punnadi[b] & R. Smith[a]

[a]Caltech Optical Observatories, California Institute of Technology, 1200 E. California Blvd., MC 11-17, Pasadena, CA, USA 91125; [b]Inter-University Centre for Astronomy & Astrophysics, Post Bag 4, Ganeshkhind, Pune, India 411 007; [c]Dunlap Institute for Astronomy and Astrophysics, University of Toronto, 50 St. George Street, Toronto M5S 3H4, Ontario, Canada

**ABSTRACT**

We have created a new autonomous laser-guide-star adaptive-optics (AO) instrument on the 60-inch (1.5-m) telescope at Palomar Observatory called Robo-AO. The instrument enables diffraction-limited resolution observing in the visible and near-infrared with the ability to observe well over one-hundred targets per night due to its fully robotic operation. Robo-AO is being used for AO surveys of targets numbering in the thousands, rapid AO imaging of transient events and long-term AO monitoring not feasible on large diameter telescope systems. We have taken advantage of cost-effective advances in deformable mirror and laser technology while engineering Robo-AO with the intention of cloning the system for other few-meter class telescopes around the world.

## 1. INTRODUCTION

Robo-AO is a new autonomous laser-guide-star adaptive-optics (AO) and science instrument currently deployed on the 60-inch telescope at Palomar Observatory. The instrument has demonstrated wavefront correction sufficient for diffraction-limited resolution imaging, 0.1"-0.25", at visible and near-infrared wavelengths over a field as large as 44" × 44". Robotic software automations[1] keep target-to-target observing overheads of < 2 minutes (including slew time) which leads to the ability to observe over one-hundred targets per night. While several initial scientific results are in press[2] and in preparation, we are continuing to exploiting this capability to execute extremely large targeted AO surveys[3,4], rapid AO imaging and characterization of transient events[5] and long-term AO monitoring[6,7] not otherwise feasible on large telescope AO systems[8]. The first of many envisioned systems is currently executing a 30-night science demonstration period at the 60-inch telescope before supporting seven competitively selected programs, over 21 nights, in the 2012B semester. Three other implementations of Robo-AO are in various stages of development: work on a clone of Robo-AO for the 2-m IUCAA Girawali Observatory near Pune, India is underway; a natural guide star only system for the 1-m Table Mountain telescope, led by Pomona College, has recently closed their AO loop on sky; and a system dedicated to precision near-infrared astrometry is being studied for deployment at the South Pole.

## 2. DESCRIPTION OF THE ROBO-AO SYSTEM

The Robo-AO system comprises several main systems: a robotic host telescope; the laser system; support electronics including the master robotic control computer; and a Cassegrain mounted adaptive optics system with integrated science instruments (see Fig. 1).

The first Robo-AO system has been deployed on the Palomar 60-inch telescope. The telescope was upgraded with full robotic control in 2004, permitting local or remote operation via TCP/IP commands[9]. In addition, the telescope uses data from a locally hosted weather station to execute automated routines to ensure telescope safety during adverse weather conditions.

The core of the Robo-AO laser system is a pulsed 10-W ultraviolet laser (JDSU Q301-HD) mounted in an enclosed projector assembly on the side of the telescope. The laser projector incorporates a redundant shutter in addition to the laser's internal shutter for additional safety; a half-wave plate to adjust the angle of projected linear polarization;

---

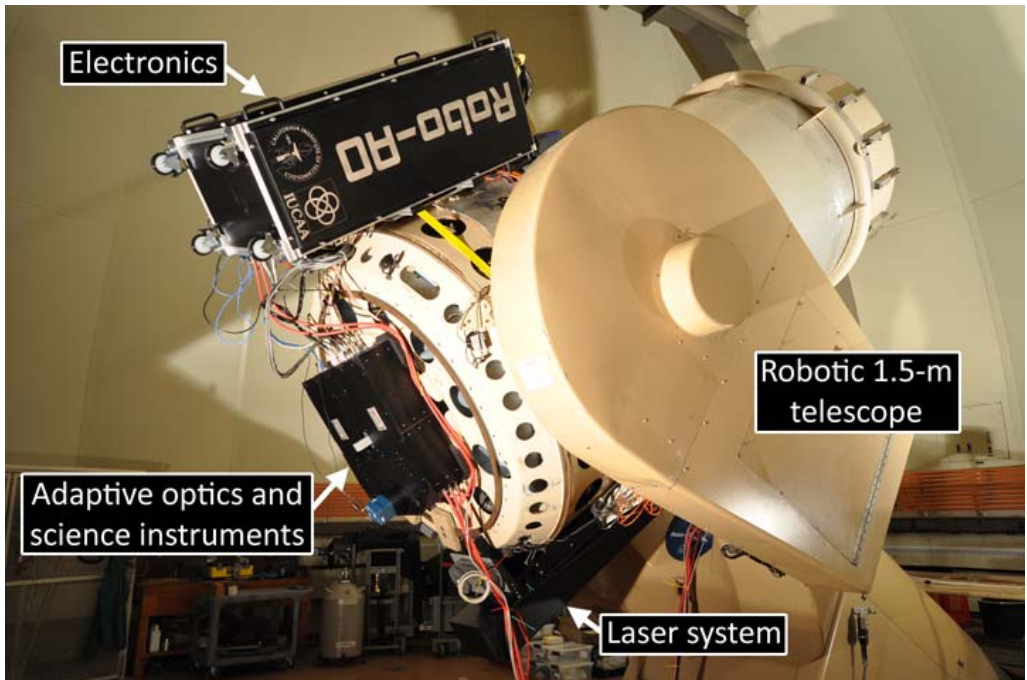

Figure 1. The Robo-AO laser adaptive-optics system. The adaptive optics and science instruments are installed at the Cassegrain focus of the robotic 60-inch telescope at Palomar Observatory. The laser system and support electronics are attached to opposite sides of the telescope tube for balance.

and an uplink tip-tilt mirror to both stabilize the apparent laser beam position on sky and to correct for telescope flexure. A bi-convex lens on an adjustable focus stage expands the laser beam to fill a 15 cm output aperture lens, which is optically conjugate to the tip-tilt mirror. The output lens focuses the laser light to a line-of-sight distance of 10 km. Within the adaptive-optics instrument, a high-speed beta-$BaB_2O_4$ Pockels cell optical shutter (FastPulse Lasermetrics[10]) is used to transmit laser light only returning from a ~400-m slice of the atmosphere around the 10 km projector focus, resulting in the laser appearing as a spot. Switching of the Pockels cell is driven by the same master clock as the pulsed laser, with a delay to account for the round trip time of the laser pulse through the atmosphere.

The ultraviolet laser has the additional benefit of being invisible to the human eye, primarily due to absorption in the cornea and lens[11]. As such, it is unable to flash-blind pilots and is considered a Class 1 laser system (i.e. incapable of producing damaging radiation levels during operation and exempt from any control measures[12]) for all possible exposures of persons in overflying aircraft, eliminating the need for human spotters located on site as normally required by the Federal Aviation Authority within the U.S.[13]. Unfortunately, the possibility for the laser to damage some satellites in low Earth orbit may exist. For this reason, it is recommended for both safety and liability concerns to coordinate laser activities with an appropriate agency (e.g., with U.S. Strategic Command within the U.S.[14]).

High-order wavefront sensing is performed with an 11×11 Shack-Hartmann wavefront sensor. The detector is an 80×80 pixel format E2V-CCD39 optimized for high quantum efficiency at the laser wavelength (71.9%) and is paired with a set of SciMeasure readout electronics. The detector is binned by a factor of 3. Each Shack-Hartmann subaperture is imaged with 2×2 binned pixels projected to a 5"×5" field of view (limited by a circular 4.8" diameter field stop). The frame transfer time is 500 μs, with frame exposures of 833 μs, corresponding to an effective frame rate of 1.2 kHz. The measured read-noise in this mode is ~6.5 electrons per pixel. The laser return signal ranges from 100 to 200 photoelectrons per subaperture per frame (depending on seeing conditions and elevation of observations), equivalent to a signal-to-noise ratio on each slope measurement of 6 to 10.

The wavefront reconstructor matrix is a synthetic modal reconstructor based on the disk harmonic functions[15] up to the 11th radial order, for a total of 75 controlled modes. The AO control loop is based on a simple integral controller with a leaky integrator to a median flat position. Individual modal gain optimization has been performed using on-sky telemetry; however it was found that an overall loop gain of 0.6 was close to optimal in a majority of situations. The closed-loop bandwidth of the system is approximately 90-100 Hz. The tip-tilt modes measured in the wavefront sensor are dominated by mechanical vibration and pointing errors. The tip-tilt signal is used to drive the laser system's uplink tip-tilt mirror, thus keeping the Shack-Hartmann pattern centered on the wavefront sensor.

The high-order wavefront corrector within Robo-AO is a micro-electro-mechanical-systems (MEMS) deformable mirror[16] (Boston Micromachines Multi-DM). Robo-AO uses 120 of the 140 actuators to adjust the illuminated surface of the mirror, sufficient in spatial resolution to accurately fit the calculated correcting shape. The actuators have a maximum surface deviation amplitude of 3.5 μm which corresponds to optical phase compensation of up to 7 μm. In typical seeing at Palomar observatory (median ~1.1"), this compensation length is greater than 5-sigma of the amplitude of the turbulence induced optical error and therefore results in significant correction headroom. Furthermore, the deformable mirror is used to compensate for static optical errors arising from the instrument and telescope at the cost of reduced dynamic range.

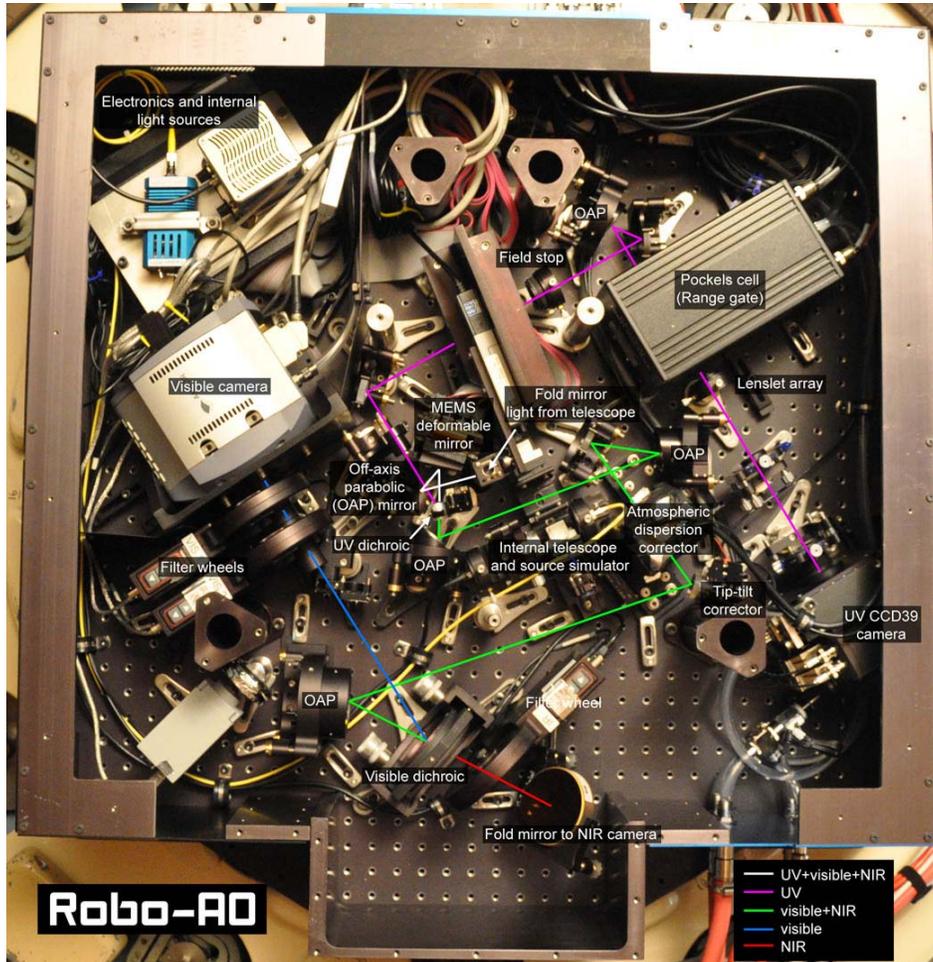

Figure 2. The Robo-AO adaptive optics system and science instruments. Light focused from the telescope secondary mirror enters through a small hole at the center of the instrument before being reflected by 90 degrees by the first fold mirror towards an off-axis parabolic (OAP) mirror. This mirror images the telescope pupil on the deformable mirror surface. After reflection from the deformable mirror, an UV dichroic splits off the laser light (violet) and directs it to the laser wavefront sensor. An additional reversed OAP mirror within the wavefront sensor corrects the non-common path optical errors introduced by the 10 km conjugate focus of the laser reflecting off of the first OAP mirror. The visible and near-infrared light (green) passing through the UV dichroic is relayed by a pair of OAP mirrors to the atmospheric dispersion corrector. The light is then reflected by the tip-tilt correcting mirror to a final OAP mirror which focuses the light towards the visible dichroic. The visible dichroic reflects the visible light (blue) to the electron-multiplying CCD and transmits the near-infrared light (red) to a fold mirror and ultimately to the infrared camera. The combined UV, visible and near-infrared light from the telescope and source simulator can be directed to the adaptive optics and science instruments by translating the first fold mirror out of the way.

Robo-AO uses four off-axis parabolic (OAP) mirrors to relay a 2' diameter full field of view from the telescope to the science cameras achromatically. The relay path includes a fast tip-tilt correcting mirror as well as an atmospheric dispersion corrector (ADC)[17] comprised of a pair of rotating prisms. At the end of the OAP relay is a visible dichroic that reflects light of λ < 950 nm to an electron-multiplying charge-coupled device camera (EMCCD; Andor iXon 888) while transmitting infrared light towards an engineering grade infrared camera (Xenics Xeva-1.7-320). The EMCCD camera has the ability to capture images with very low electronic (detector) noise[18,19], at a frame rate which reduces the intra-exposure image motion to below the diffraction-limited angular resolution. By re-centering and stacking a series of these images, a long-exposure image can be synthesized with minimal noise penalty. The EMCCD camera can also be used to stabilize image motion on the infrared camera; measurements of the position of an imaged astronomical source can be used to continuously command the fast tip-tilt to re-point the image to a desired location. Ahead of each camera is a set of filter wheels with an appropriate set of astronomical filters.

An internal telescope and source simulator is integrated into the Robo-AO system as a calibration tool. It can simultaneously simulate the ultraviolet laser focus at 10 km and a blackbody source at infinity, matching the host telescope's focal ratio and exit pupil position. The first fold mirror within Robo-AO directs all light from the telescope's secondary mirror to the adaptive-optics system. The fold mirror is also mounted on a motorized stage which can be translated out of the way to reveal the internal telescope and source simulator.

## 3. ON-SKY ADAPTIVE OPTICS CORRECTION

Robo-AO offers several distinct LGS observation modes, targeting different areas of wavelength and sky coverage parameter space, distinguished by use of either an on-axis science target or a nearby field star for tip-tilt wavefront sensing. Currently, we achieve the best visible-light Robo-AO performance by using post-facto shift-and-add image processing algorithms that mitigate residual atmospheric tip-tilt errors. This requires a source brighter than an in-band magnitude of $m \sim 13.5$ for effective post-processing. The future low-noise wide-field infrared camera (described in section 5.1) will allow Robo-AO to observe in the visible at the diffraction limit using much fainter tip-tilt stars. When estimating tip-tilt from the science target using the infrared camera, Robo-AO will be able to achieve diffraction-limited visible-light performance for point sources as faint as $m_V = 17$ in median conditions (Table 1 presents detailed performance calculations). When using an off-axis near-infrared field star for tip-tilt and assuming the minimum field of the H2-RG camera, approximately 30% sky coverage can be accessed at the cost of up to an additional 0.23" of RMS tip-tilt (two-axis) error to be added in quadrature. For 90% sky coverage, there is an additional (up to) 0.35" of RMS tip-tilt error, providing the equivalent of exquisite seeing over nearly the entire Northern sky.

We have maintained detailed error budgets for the expected adaptive optics performance of Robo-AO under different observing conditions (see Table 1). The error budgets use measured $C_n^2(h)$ profiles from an onsite MASS-DIMM atmospheric turbulence monitor collected over a year. These error budgets are based on our team's considerable real-world experience building and commissioning AO systems at Palomar and MMT, and in measuring the performance of the Keck AO system. At Palomar, we find delivered LGS Strehl performance is usually within 10% (relative) of our detailed model predictions, giving us strong confidence in our performance predictions for Robo-AO. We expect dramatic image improvement in all conditions at infrared wavelengths, and access to visible near-diffraction-limited imagery under median conditions. During 25th best percentile seeing, Robo-AO achieves appreciable Strehl ratios in the visible red wavelengths operating up to 40° in zenith angle. As seeing conditions degrade, the available sky fraction for visible correction decreases. However, even under poor conditions (75th percentile seeing), very reasonable correction in the NIR can still be achieved within 10° of zenith. For guide sources brighter than $m_V=15$, H-band Strehl ratio performance is limited by high-order wavefront errors from the laser system. Beyond $m_V=17$, tip-tilt errors start to dominate, affecting the overall AO correction.

Table 1. Example error budget for Robo-AO under different seeing conditions ($r_0$) and for different zenith angles (z) assuming an on-axis $m_V$=17 star for tip-tilt sensing and on-axis science target. Measurement error arises from finite laser photoreturn, WFS read noise, sky noise, dark current, and other factors. Focal anisoplanatism is an error arising from the finite altitude of the Rayleigh LGS resulting in imperfect atmospheric sampling. Multispectral error arises from differential refraction of UV and visible/NIR rays and grows rapidly for large off-zenith angles. For extended targets an additional 65 nm RMS of angular anisoplanatism error would be present at a field position 10 arcseconds removed from the direction of the LGS (finite aperture effects on the 1.5 m telescope mitigate the classical calculation of isoplanatic angle). 'mas' indicates milli arc seconds, " indicates arc seconds.

| | Percentile Seeing | 10% | 25% | 50% | 75% |
|---|---|---|---|---|---|
| | | 0.67" @ z=0° | 1.02" @ z=40° | 1.12" @ z=20° | 1.69" @ z=10° |
| | | $r_0$ = 15 cm | $r_0$ = 9.8 cm | $r_0$ = 8.9 cm | $r_0$ = 5.9 cm |
| **High-order Errors** | | \multicolumn{4}{c}{Wavefront Error (nm)} | | | |
| Atmospheric Fitting Error | | 39 | 56 | 61 | 85 |
| Bandwidth Error | | 47 | 52 | 65 | 92 |
| High-order Measurement Error | | 41 | 46 | 57 | 81 |
| LGS Focal Anisoplanatism Error | | 60 | 102 | 96 | 131 |
| Multispectral Error | | 0 | 74 | 11 | 3 |
| Scintillation Error | | 13 | 26 | 22 | 29 |
| WFS Scintillation Error | | 10 | 10 | 10 | 10 |
| Uncorrectable Tel / AO / Instr Aberrations | | 38 | 38 | 38 | 38 |
| Zero-Point Calibration Errors | | 34 | 34 | 34 | 34 |
| Pupil Registration Errors | | 21 | 21 | 21 | 21 |
| High-Order Aliasing Error | | 13 | 19 | 20 | 28 |
| DM Stroke / Digitization Errors | | 5 | 5 | 5 | 5 |
| **Total High Order Wavefront Error** | | **112 nm** | **168 nm** | **156 nm** | **219 nm** |
| **Tip-Tilt Errors** | | \multicolumn{4}{c}{Angular Error (mas)} | | | |
| Tilt Measurement Error | | 19 | 24 | 27 | 39 |
| Tilt Bandwidth Error | | 14 | 18 | 21 | 30 |
| Science Instrument Mechanical Drift | | 6 | 6 | 6 | 6 |
| Residual Telescope Pointing Jitter | | 2 | 3 | 2 | 2 |
| Residual Centroid Anisoplanatism | | 1 | 2 | 2 | 2 |
| Residual Atmospheric Dispersion | | 0 | 1 | 0 | 0 |
| **Total Tip/Tilt Error (one-axis)** | | **24 mas** | **30 mas** | **35 mas** | **50 mas** |
| **Total Effective Wavefront Error** | | **137 nm** | **176 nm** | **185 nm** | **241 nm** |
| Spectral Band | λ | λ/D | Strehl / FWHM | Strehl / FWHM | Strehl / FWHM | Strehl / FWHM |
| r' | 0.62 μ | 0.08" | 17% / 0.09" | 5% / 0.14" | 4% / 0.15" | 0% / 0.59" |
| i' | 0.75 μ | 0.10" | 28% / 0.10" | 12% / 0.15" | 10% / 0.15" | 2% / 0.21" |
| H | 1.64 μ | 0.22" | 75% / 0.22" | 63% / 0.23" | 59% / 0.24" | 38% / 0.26" |
| K | 2.20 μ | 0.30" | 86% / 0.30" | 78% / 0.30" | 76% / 0.30" | 62% / 0.30" |

As validation of our high-order wavefront error budget, we have captured fast frame rate visible data (≥ 30Hz) and simulated a long exposure by using post-facto tip-tilt correction on the recorded data (this method of visible-only observing is empirically limited to targets of SDSS band magnitudes of ~13.5 or brighter). Our best post-facto tip-tilt corrected stellar image during early commissioning in 2011, was of 61 Cyg A, with a Strehl ratio of 14% in r'-band (λ=620 nm), corresponding to a residual high-order wavefront error of 140±9 nm RMS and 0.10" FWHM. This was taken during a time of less-than-median seeing, ~1.3". During typical observing (e.g. Figs. 3, 5 and 6), where objects are generally scattered around the sky, we generally obtain residual wavefront errors in the 160 to 200 nm RMS range as indicated by the image analysis of Fig. 4 (see Table 2). This level of wavefront correction leads to a modest ability to detect and characterize stellar companions at visible wavelengths (see Fig. 4).

Table 2. Measured image quality metrics from the upper star in Figure 3. The brightest star in the field (leftmost) was used for post-facto image registration.

| λ (nm) | FWHM | Strehl | RMS WFE (nm) |
|---|---|---|---|
| 625 | 0.11" | 7.1% | 161 |
| 765 | 0.12" | 14.7% | 168 |
| 890 | 0.14" | 20.4% | 178 |

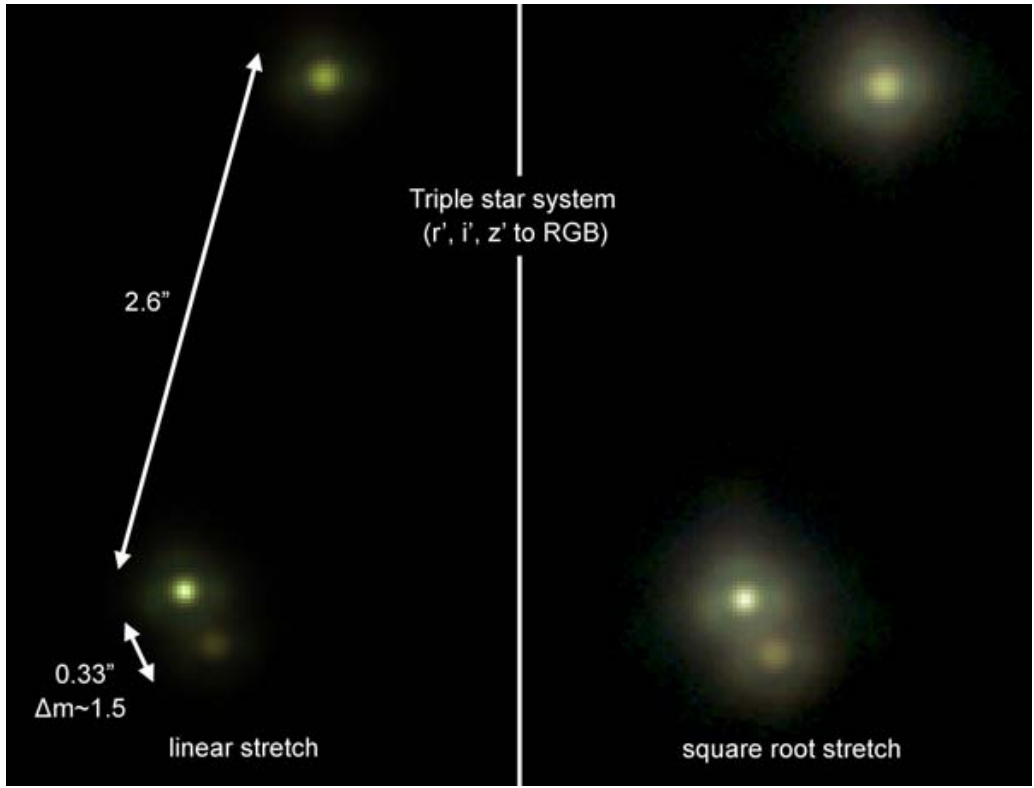

Figure 3: Robo-AO observation of a triple star system in r'-, i'- and z'-band.

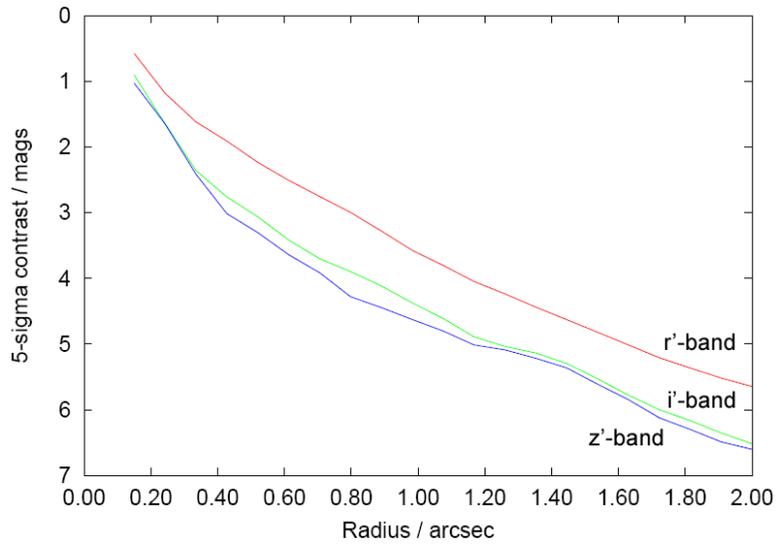

Figure 4. Measured 5-sigma contrast ratio versus field separation on a typical observation of a target brighter than m~13.5. This allows for the detection and characterization of binary and multiple systems (e.g. Fig. 6).

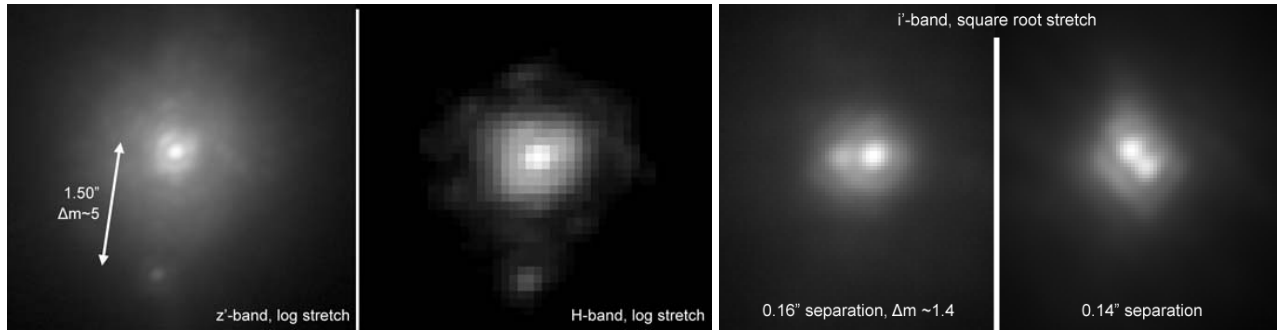

Figure 5. Binary star systems imaged in the visible and near-infrared by Robo-AO.

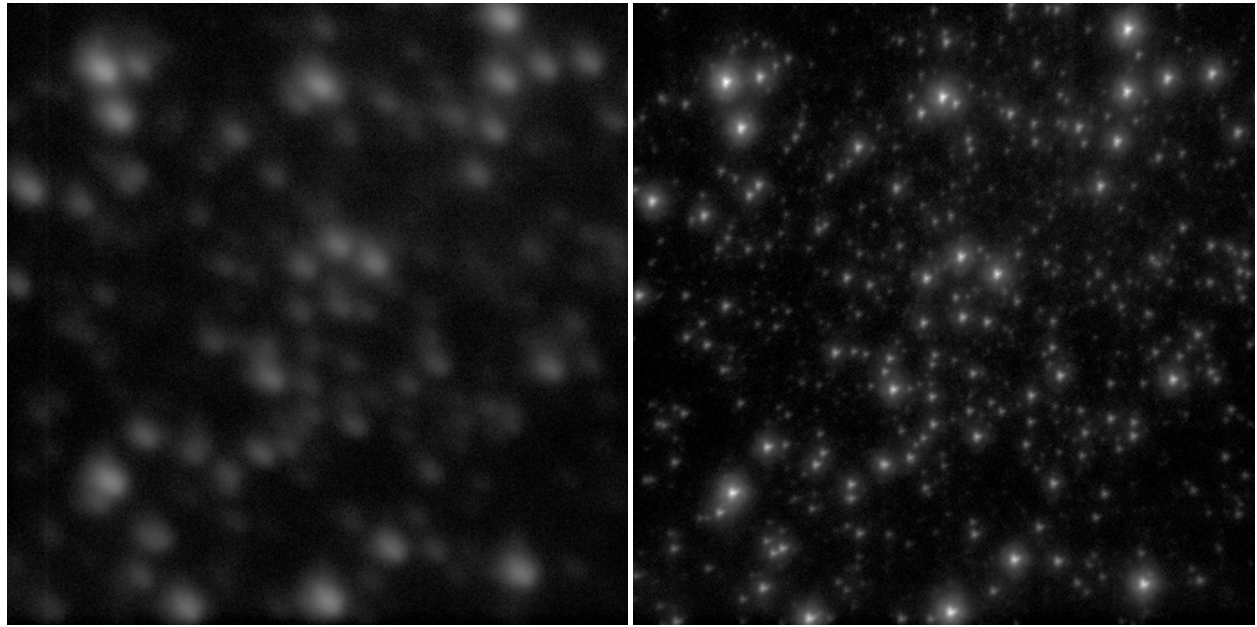

Figure 6. Left: A 44" × 44" field-of-view, 2-minute long seeing-limited image of the core of the globular cluster Messier 3 in z-band (λ = 830 – 950 nm). The same image shown with Robo-AO adaptive optics correction.

## 4. EDUCATIONAL AND SYNERGISTIC ACTIVITIES

### 4.1 Student education

The Robo-AO collaboration is specifically interested in targeting undergraduate education to encourage promising students to continue into graduate programs emphasizing science and technology. In our group, students learn through inquiry, helping senior researchers identify questions fundamental to both the astronomy and the engineering that we undertake. Student participation is organized around research projects that they help identify and then complete.

Robo-AO has already involved 10 undergraduates through Caltech's Summer/Minority Undergraduate Research Fellowship (SURF/MURF) program which supports undergraduates for ten weeks of guided research alongside a diverse peer group. Our students have made major contributions to the Robo-AO project. As an example, during the summer of 2011, two students worked separately on an intelligent queue scheduler and real-time telemetry analysis and display tools respectively; both of their projects were tested by the students at the P60 during a Robo-AO commissioning run in August 2011. This summer, students are developing software to provide photometric quantities, image quality and companion detection for fast evaluation; evaluating the limitations of HgCdTe detectors for precision astrometry which will be later used in the new Robo-AO infrared camera; and evaluating the current astrometric capabilities of Robo-AO by analyzing hours of visible-light observations of star clusters.

### 4.2 Synergistic projects

We have established an educational collaboration with Pomona College which has proven to be particularly fruitful. This collaboration inspired the creation of a new AO lab at Pomona aimed at hands-on education of undergraduates, and the deployment of a natural-guide-star version of Robo-AO on their 1-m telescope at Table Mountain. As part of the ongoing collaboration, the Robo-AO has supplied the Pomona AO system with the full real-time Robo-AO software engine. The prototype Pomona system, dubbed 'KAPAO-Alpha', which demonstrates the basic operation of a natural guide star adaptive optics system has recently closed the AO loop on-sky and their results have recently been reported[20]. The final KAPAO system (no 'Alpha'), which will be a remote-access system offering simultaneous dual-band, diffraction-limited imaging at visible and near-infrared wavelengths, delivering an order-of-magnitude improvement in point source sensitivity and angular resolution relative to the current seeing limits is in development and is expected to be deployed in the near future[21]. Beyond the expanded scientific capabilities enabled by AO-enhanced resolution and sensitivity, the interdisciplinary nature of the instrument development effort provides an exceptional opportunity to train a broad range of undergraduate STEM students in AO technologies and techniques. The Pomona Table Mountain AO system is an exciting example of how Robo-AO can expand the interaction between research and learning through adaptive optics.

### 4.3 Outreach

The Robo-AO collaboration has published a web site detailing the goals, science and technical details of the project, geared towards accessibility to the general public: http://www.astro.caltech.edu/Robo-AO/. Updates are provided, highlighting major project milestones.

Dissemination of results to the broader public, including interpretation of science discovery and its impact on society, are made to the public by Palomar Observatory's Visitor Center (http://www.astro.caltech.edu/palomar/exhibits/) which draws over 100,000 visitors annually. A web-based kiosk presentation highlighting adaptive optics science is on display in the Visitor Center and available on the web. The observatory has also embraced other online social media, and the results Robo-AO commissioning have appeared on Palomar Observatory's blog and Facebook page.

In addition, Robo-AO has recently been augmented with an eyepiece for the direct viewing of visible-light adaptive optics corrected astronomical objects (Fig. 7). The eyepiece is fed by a selectable visible beam splitting mirror mounted on a removable kinematic base between the visible dichroic and visible filter wheels. This allows for simultaneous imaging with the visible camera. While access is currently limited, we intend to support future eyepiece observing by visiting astronomy classes from Caltech and Pomona College.

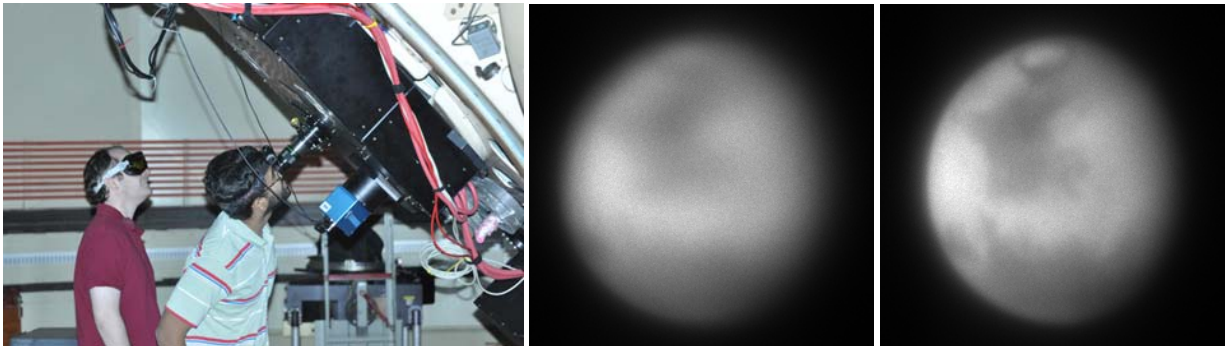

Figure 7. Members of the Robo-AO team observing Mars with the eyepiece attached to Robo-AO (left). Simultaneous i-band images of Mars captured without adaptive optics correction (middle) and with the Robo-AO system running (right).

## 5. FUTURE WORK

### 5.1 Infrared camera upgrade

The Robo-AO collaboration is currently in the process of upgrading the infrared imaging system from an engineering grade camera to a low-noise wide-field imager using a 2.5 μm cutoff Teledyne HAWAII-2RG™/HgCdTe detector (H2-RG). The science grade H2-RG detector has ~ 0.01 $e^-$/sec dark current, < 5 $e^-$ readout noise[22] and slightly higher quantum efficiency than the InGaAs, offering 45-80 times more sensitivity (Table 3). The new camera will consist of a simple reimaging system with cold stop and filters (JHK, H-br, CH4 on & off, a blank, and a spare) at the internal pupil plane. The H2-RG detector (on order and to be delivered) has a guaranteed useable area of 780×780 pixels. We have adopted a plate scale of 0.086"/pixel, Nyquist at J-band, corresponding to a 67"×67" field, although we will transmit the full 2' field to the detector if a larger useable area of the detector is delivered.

Table 3. Optical format and infrared sensitivity of the Robo-AO InGaAs and H2-RG infrared cameras.

| Detector | Field of view | Plate Scale | J sensitivity[1] | H sensitivity[1] | K ($\lambda$=2.2 um) sensitivity[1] |
|---|---|---|---|---|---|
| InGaAs | 32.0"×25.6" | 0.10"/pixel | 14.9 | 15.2 | N/A |
| H2-RG | > 67"×67" | < 0.086"pixel | 19.7 | 19.4 | 18.2 |

[1]Faintest magnitude to reach a signal-to-noise ratio of 10 in 120 s of exposure. These assume a residual wavefront error of 185 nm RMS, expected AO performance under median seeing conditions (table 1).

The Robo-AO instrument has an external camera port (Fig. 8) designed to accommodate an instrument matched to the expected 70-kg mass of the camera.

Furthermore, as part of our integration plan, we will develop automated routines for configuring the high-speed NIR tip-tilt sensing and recording of NIR science data in much the same way as has been done with the visible Robo-AO camera. Special procedures for dealing with infrared data, including taking sky-flats, backgrounds, and darks, as well as compensating for nonlinear responses and amplifier glow, will be developed and included as part of a standard data pipeline. We will also maintain a flexible FITS header for all data captured with Robo-AO's NIR and visible cameras, making it Virtual Observatory-compliant as recommended in the Astro-2010 Decadal survey[23] for possible later public release and archiving.

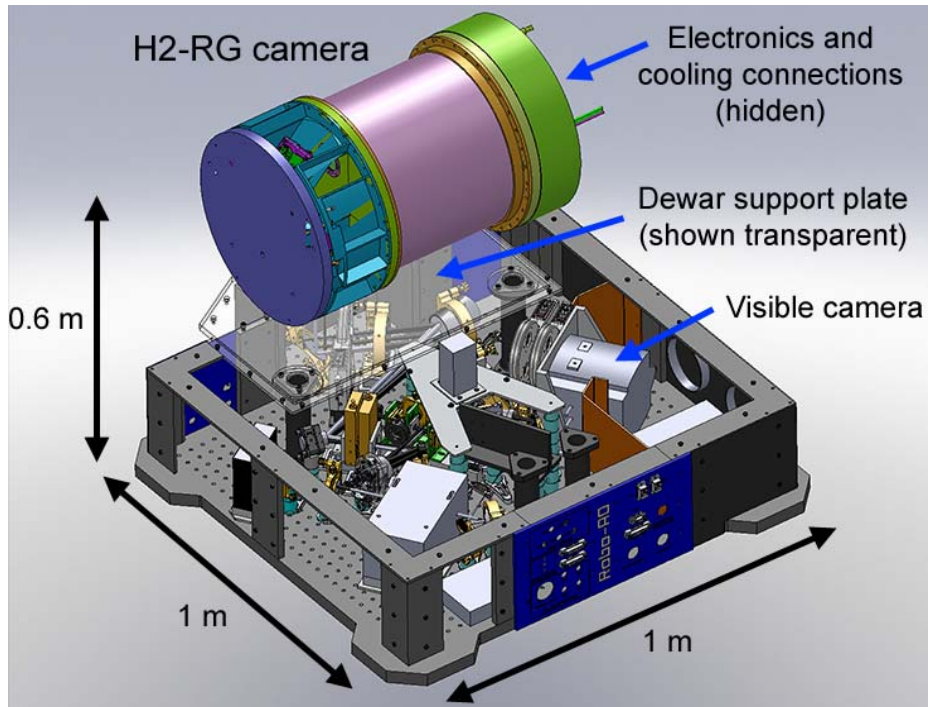

Figure 8. Solidworks model of the as-built Robo-AO system with a model of the Keck infrared tip-tilt sensor (similar in function, size and weight to the planned Robo-AO NIR camera) attached.

### 5.2 Clone of Robo-AO for the IUCAA Girawali Observatory

A clone of the current Robo-AO system is currently being developed for the 2-m IUCAA Girawali Observatory telescope[24] in Maharashtra, India. Adaptations in optical prescriptions are being made to accommodate the slightly different F/#, F/10 vs. F/8.75, different mounting interface and telescope interface software. No other major changes are anticipated in an effort to minimize any further development and control costs. The system is expected to see first light in 2013.

### 5.3 Robo-AO at the South Pole

The autonomous function of Robo-AO is particularly well suited for remote siting. In particular, Dekany[25] has proposed the development of a 2.4m-class Robo-AO telescope dedicated to the unique infrared and astrometric advantages of the United States Amundsen-Scott Station at the South Pole. The ability of Robo-AO to mitigate seeing up to 2" FWHM, proven at Palomar, confirms the suitability of Robo-AO to mitigate the boundary layer at South Pole, allowing the first generation of 2m-class Antarctic telescope to exploit the logistical, geographic, and winter staffing advantages of the South Pole.

## AKNOWLEDGEMENTS

The Robo-AO system is supported by collaborating partner institutions, the California Institute of Technology and the Inter-University Centre for Astronomy and Astrophysics, by the National Science Foundation under Grant Nos. AST-0906060 and AST-0960343, by a grant from the Mt.Cuba Astronomical Foundation and by a gift from Samuel Oschin. The infrared camera upgrade is additionally supported in part by the National Science Foundation under Grant No. AST-1207891 and by the Office of Naval Research under grant N00014-11-1-0903. We are grateful for the continued support of the Palomar Observatory staff for their ongoing support of Robo-AO on the 60-inch telescope, particularly S. Kunsman, M. Doyle, J. Henning, R. Walters, G. Van Idsinga, B. Baker, K. Dunscombe and D. Roderick.